\documentclass[10pt,sigplan,nonacm]{acmart}
\usepackage{amsmath}
\usepackage{xcolor}
\usepackage{listings}
\usepackage{comment}
\usepackage{placeins} 

\lstdefinestyle{py}{
  language=Python,
  basicstyle=\ttfamily\small,
  keywordstyle=\color{blue!60!black},
  commentstyle=\color{green!40!black},
  stringstyle=\color{orange!60!black},
  showstringspaces=false,
  breaklines=true,
  frame=single,
}

\AtBeginDocument{%
  }

\setcopyright{none}                
\renewcommand\footnotetextcopyrightpermission[1]{} 
\usepackage{xspace}

\acmISBN{}

\begin{document}

\newcommand{\sys}{\textsc{Plainbook}\xspace}
\newcommand{\mypara}[1]{\smallskip\noindent\textbf{#1.}}
\newcommand{\mynote}[1]{\textbf{\textsf{\textcolor{red}{#1}}}}
\newcommand{\ie}{\emph{i.e.,}\xspace}
\newcommand{\eg}{\emph{e.g.,}\xspace}
\newcommand{\codeidx}{\mathit{i}_{\mathit{code}}}
\newcommand{\outidx}{\mathit{i}_{\mathit{out}}}
\newcommand{\execidx}{\mathit{i}_{\mathit{exec}}}
\newcommand{\set}[1]{\{#1\}}
\newcommand{\setup}{\mathit{setup}}
\newcommand{\test}{\mathit{test}}

\title{{Plainbook}: Data Science, in Plain Language}

\author{Luca de Alfaro}
\email{luca@ucsc.edu}
\orcid{0000-0003-3856-4576}
\affiliation{%
\institution{University of California}
\city{Santa Cruz}
\state{California}
\country{USA}
}

\author{Mathis Aubert}
\email{maubert@ucsc.edu}
\orcid{0009-0008-3412-1296}
\affiliation{%
\institution{University of California}
\city{Santa Cruz}
\state{California}
\country{USA}
}

\author{Ranjit Jhala}
\email{rjhala@ucsd.edu}
\orcid{0000-0002-1802-9421}
\affiliation{%
\institution{University of California}
\city{San Diego}
\state{California}
\country{USA}
}

\author{Eliana Pastor}
\email{eliana.pastor@polito.it}
\orcid{0000-0002-3664-4137}
\affiliation{%
\institution{Politecnico di Torino}
\city{Torino}
\country{Italy}
}

\author{Elena Baralis}
\email{eliana.pastor@polito.it}
\orcid{0000-0001-9231-467X}
\affiliation{%
\institution{Politecnico di Torino}
\city{Torino}
\country{Italy}
}


\begin{abstract}
Jupyter Notebooks have become widely adopted in data science, as they allow the sharing of reproducible computational analysis.
They are, however, accessible only to people who understand computer code.
To reach the broader audience of scientists interested in data analysis and computation, but unfamiliar with code, we introduce \sys, notebooks centered on natural language rather than code.

\sys is based on two principles: \emph{promote the natural language descriptions}, and \emph{verify the values}.
In \sys, the natural language descriptions are preserved, rather than the resulting code; the code is generated automatically from the cell descriptions.
As natural language is read top to bottom, \sys adopts a linear execution semantics, in which cells are guaranteed to be executed in the order in which they appear; there is no ``hidden state'' or out-of-order execution as in Jupyter.
To allow users who may not understand code to verify the correctness of the computation, we have built into \sys verification mechanisms centered on \emph{values} and value inspection.
These include mechanisms that focus on individual cells, akin to unit tests, as well as global mechanisms.
Both the linear execution semantics, and the verification mechanisms, are underpinned by a \emph{snapshot kernel} that caches execution states and makes execution and verification efficient.
\end{abstract}



\keywords{Jupyter Notebooks, Data Science, AI Generated Code, Coding in Natural Language, AI Assisted Programming, Conversational Notebooks, Computational Notebooks, Linear Execution Semantics, Unit Testing, Reproducible Research, Software Verification.}


\maketitle

\settopmatter{printfolios=true}
\pagestyle{plain}                

\section{Introduction}

Jupyter Notebooks have become very widely adopted in data science,
because they allow users to create and share analysis and results
in a way that is \emph{verifiable}, \emph{reproducible}, and
\emph{extensible} \cite{kluyverJupyterNotebooksaPublishing2016,pimentelUnderstandingImprovingQuality2021,pertsevaTheoryScientificProgramming2024,huangHowScientistsUse2025a}.
A Jupyter Notebook consists of computer code (most commonly Python,
but also other languages such as R, Julia, and SQL), along with the
output created by the code \cite{perezIPythonSystemInteractive2007}.
The output includes text output and also graphics, such as plots,
diagrams, maps, and other visualizations that enable
researchers and data scientists to include, in a single document,
\emph{both} the code that produced the results and the results
themselves.
The complete analysis can be shared in a way that makes it easy
to \emph{reproduce} and \emph{verify} \cite{Wangetal2019}.
Anyone can read the code and run it to reproduce the results and check
that the results included in the notebook are indeed generated
by the code performing the analysis.
Prior to notebooks, users had to keep extra metadata to associate
the results with the analysis and code that produced them. Notebooks
make this connection explicit: one can simply ``re-run'' the notebook
to reproduce and validate the results.
Further, when a notebook and its underlying data is shared,
recipients can see how the results are obtained, and can
modify the code to perform additional analyses or alter the
settings of the analysis in the notebook.

The above is only true if the notebook authors and recipients can \emph{read} computer code.
Data science has become very widely used, and its results are often shared with people who are not computer scientists.
Furthermore, since AI has become quite proficient in generating code from natural language prompts,
notebooks are often created by people who \emph{cannot write} computer code.

For example, consider the recent experience of one of the authors, who was serving on a central university committee that frequently needed to analyze data.
To generate the Jupyter notebooks, the author wrote precise descriptions in natural language of the computation to be done in each cell, and prompted the AI to generate the code; the code was then run to generate the results.
The author was the \emph{only} computer scientist on the committee.
The other committee members could look at the results, but to them the code was so much
mumbo-jumbo, and so the verifiability and extensibility advantages of Jupyter notebooks
were utterly lost to them.
Had the natural language descriptions of the cells, rather than the code, been preserved in the notebooks, the notebooks would have been accessible to all.

In short, recent developments in AI are ushering in a world where computer code is generated via natural language, and computer code is becoming a lower-level language, necessary for program execution but of lesser interest to developers.
%
%
Thus, the central question becomes: how can we evolve data science notebooks so that the broader audience of non-programmers (or rather, \emph{natural language} programmers) can benefit from their key strengths of verifiability, reproducibility, and extensibility?
We posit that this question can be answered by exploiting the following two observations.

\mypara{Promote the Descriptions}
If the natural language descriptions used to create the cells had been preserved and centered as the primary content, the notebooks would have been far more accessible to the non-programmers on the university committee, enabling them to \emph{modify the descriptions} to regenerate the code to satisfy their analysis needs.
Of course, natural language is not quite another programming language.
Despite the remarkable advances in AI, the sheer ambiguity of natural language, alongside the residual limitations of AI, make translation from  natural language prompts to computer code a process where errors or misunderstandings can (and do!) easily creep in.


\mypara{Verify the Values}
For people to have confidence in the results produced by natural language notebooks, we require mechanisms
that let the user systematically \emph{verify} the intermediate values computed \emph{locally} within individual cells, and \emph{globally} across multiple stages of analysis.
By focusing verification on the values, we can give the user confidence that they have implemented their intended analysis, without requiring them to inspect the generated code itself.

%
\smallskip
We used these insights to design \sys,
a data science notebook designed to
bring verifiability, reproducibility,
and extensibility to non-programmers.
Specifically, \sys is organized around
the following key design decisions that implement
the above observations.

\smallskip
\noindent
\emph{1. Natural language descriptions.}
\sys does not discard the natural language descriptions used to generate cells; instead, it preserves them and presents them as the primary material of that cell.

\smallskip
\noindent
\emph{2. Stable implementation.}
The code is generated from the cell descriptions via AI. 
To achieve predictable behavior, we save the code alongside the descriptions, and we ask AI to modify the code only if necessary to reflect changes in the descriptions. 

\smallskip
\noindent
\emph{3. Linear execution model.}
\sys adopts a strictly linear top-to-bottom execution model. 
Each cell is executed in the state resulting from the previous cell's execution.
In particular, executing a cell twice is idempotent.
This matches the way natural language works: reading a paragraph twice does not change its meaning.
This also eliminates the problem of ``hidden state'' in Jupyter notebooks, where the state depends on the order in which cells have been executed \cite{Wangetal2019,huangHowScientistsUse2025a}.
Eliminating hidden state is beneficial when working with natural language, since there is no convenient way to examine variable values.

To support this execution model, we have developed a \emph{check-pointing kernel} that stores,
and makes available for analysis, the checkpoints (the states or data values) \emph{before}
and \emph{after} the execution of each cell, and enables notebook cells to
be edited and re-run efficiently from their before-states. 

\smallskip
\noindent
\emph{4. Value verification.}
Relying on the check-pointing kernel,
\sys implements new ways to verify values:
\begin{itemize}
\item \emph{cell verification} lets the
user use AI to verify that the code in the cell
implements its natural language description,

\item \emph{cell tests} let users run a particular cell using data that is modified and often simplified per their instructions, \eg
to allow users to test cell behavior in scenarios of interest and under conditions that are conducive to user inspection,

\item \emph{global tests} let the user
verify relationships about the values of states
spanning multiple points in the notebook, \eg to
check that all data having some characteristics
is still present in the notebook results after some
filtering.

\end{itemize}
These verification methods can be used both by
the user who creates a \sys, and by users with
whom the \sys has been shared, thus implementing
the central goal of allowing \emph{verifiable} sharing.
Thus, for instance, a user can verify that a cell
or a whole notebook implements its natural language
description; further, the user can do so using an AI
of their choice.

In the following, after a review of related work,
we first describe the notebook structure and
its execution model.
We then describe in detail the natural-language based verification tools we have implemented in \sys.
Lastly, we describe the implementation, including what information is passed to AI for code generation, how the check-pointing kernel is implemented, and how we ensure that the code and natural language portions remain
consistent as the notebook is edited.


\section{Related Work}

\begin{figure*}[!t]
  \centering
  \includegraphics[width=\textwidth]{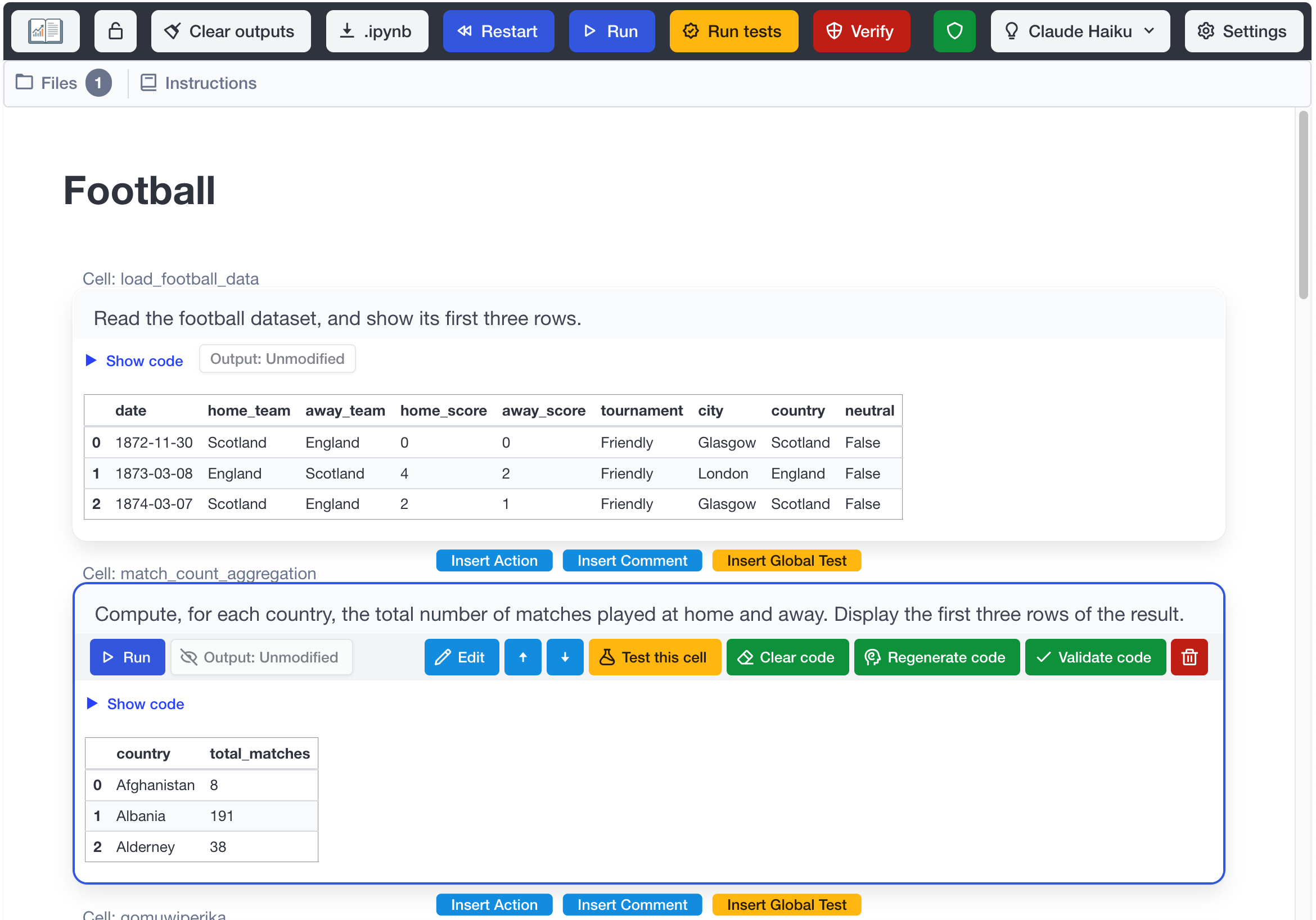}
  \caption{Top bar, and two action cell in \sys.
  The top bar allows users to select the AI model in use, restarting and running all the cells, and using AI to verify that the content of the \sys is safe.
  Below the top bar is a bar with a tab for files, and one for general instructions. 
  The \emph{Files} tab is used to select the files that are used by the notebook (in this case, the football csv dataset), so that AI knows how to access them when generating the code.
  The first two cells of the \sys are shown; the second cell is focused, so that the actions available to the user are shown. 
  Users can run the cell, validate it via unit tests (see Section~\ref{sec-unit-tests}), clear the code and regenerating it (possibly with a different AI model), and validate the code with respect to the description (see Section~\ref{sec-cell-verification}).
  The dataset is from \cite{Jurisoo2024}.
  }
  \Description{Screenshot of \sys top bar, and an action cell.
  The top bar includes buttons Restart, Run, Run tests, Verify, a button to change the AI provider used (Claude Sonnet is used), and a button to access the settings.
  Then, a cell follows, with description: Read the football dataset, and show its first three rows.  The first rows of an international soccer matches dataset follow, with columns date, home-team, away-team, home-score, away-score, tournament, city, country, and neutral.}
  \label{fig-teaser}
\end{figure*}

Translating from natural language to code, and more generally, generating code from natural language interactions, is one of the main uses of large AI models \cite{chenEvaluatingLargeLanguage2021,roziereCodeLlamaOpen2023,zhengSurveyLargeLanguage2023,jiangSurveyLargeLanguage2026,houLargeLanguageModels2024}.
This use of AI has become so immensely successful that multi-billion dollar companies have made it one of their main lines of business, if not the only one: think of Anthropic with Claude \cite{anthropic2024claude}, Google with Gemini \cite{geminiteam2023gemini}, OpenAI with Codex \cite{chenEvaluatingLargeLanguage2021},  and Microsoft with Copilot \cite{github2026copilot}.
Our work here certainly breaks no new ground in this respect, and in fact, \sys relies on the above AI models for code generation.
The novel aspects of \sys are implemented on top of the code-generating abilities of the above AI models.

The idea of extending computational notebooks to \emph{conversational} ones, in which computation descriptions augment and supplement code, has been advocated in \cite{weberComputationalConversationalNotebooks2024}.
That work details a broad spectrum of approaches for generating notebook code via interaction with AI agents, and mentions that the on-demand execution semantics of Jupyter notebooks may not mesh well with user interaction based on natural language. 
In this work, we follow the direction advocated by \cite{weberComputationalConversationalNotebooks2024}, and we focus on a specific approach, in which the notebook code is generated for descriptions of what each cell should do. 
We implement the linear execution semantics, similarly to Marimo \cite{Marimo2023,Gyarmatietal2025}, and we explore how users can test notebooks using entirely natural language.

Google Colab offers both a parallel and a contrasting example \cite{GoogleColab,Bisong2019}.
Similarly to \sys, in Colab it is possible to create a new cell using AI.
Users click to add the cell, then click to ask for AI help.
They enter a prompt, and from the prompt, the code is produced.
Users are then expected to inspect the code and accept it, and once they do, the cell persists the code --- but not the original prompt.
People who want to verify or modify the cell can ask AI to explain the cell content, and modify it.
But AI, when explaining the cell code, generates a very detailed description of how the code \emph{works}, rather than of what the code \emph{does} at a high level. Hence, the AI-generated description loses the conciseness of the original high-level prompt.
AI in Colab also allows users to easily modify the behavior of cells, but naturally the problem is that the original, high-level natural language description of the \emph{current} behavior of a cell is not there.
The whole idea of \sys is to make the natural language, rather than the code, what is persisted, shown, and edited, and to make the computer code the behind-the-scenes means for execution.
This change of perspective has deep consequences for how users can generate the code, and validate that the (unseen) code responds to their specifications.

In Claude Code, Gemini, Codex, Cursor \cite{cursor2026,he2026cursor}, and similar environments, users can develop code starting from natural language descriptions.
Even users unfamiliar with code can develop sophisticated systems in such a way, in a process that has become known as \emph{vibe-coding} \cite{Meskeetal2025,Fawzyetal2025}.
Again, the key difference with respect to \sys is that the product of such tools is the code: the natural language is not present.
It is true that Claude Code, Codex, Cursor, and the other tools can generate detailed and beautifully written explanations of the work done, but this is not equivalent to retaining the cell descriptions. 
In such tools, code is generated iteratively via a sequence of prompts, which incrementally specify the computation and fix any problems with the implementation. 
In contrast, in \sys the code is generated from a single, comprehensive natural language description for each cell. 
This natural language specification persists, and the code can at any time be recreated from it; the tests are also in natural language and can be used against any newly generated code.
Another consequence of the persistence of natural language is that when a \sys is shared, the recipient can directly read the natural-language description of the computation, and then use AI to check that the code implements it.

\emph{Literate programming} was proposed by Knuth as a way to inter-twine the natural language description of \emph{what} the code does, and \emph{how} the code does it \cite{knuthLiterateProgramming1984}.
In spirit, what it does is close to what we aim to do.
Of course, at the time when literate programming was created, there was no way to generate code from natural language, and thus it is the natural language itself that contains code primitives as inserts.

Jupyter notebooks are the direct inspiration for \sys \cite{perezIPythonSystemInteractive2007,kluyverJupyterNotebooksaPublishing2016}.
Jupyter notebooks have had an enormous impact in the way in which data science and research in computing is done, as they allow for the joint presentation of results, and of the code that generated them \cite{randlesUsingJupyterNotebook2017,pimentelUnderstandingImprovingQuality2021,huangHowScientistsUse2025a}.
The joint presentation makes it possible for anyone to reproduce the results, and thus validate them.
As the original code is present, it can be easily modified to experiment with variations of the analysis, or extended to perform additional tasks.
Many environments enable the AI-assisted creation of Jupyter notebooks, among which Colab.

In a Jupyter notebook, each cell can be executed on demand, and the same cell can be executed multiple times.
While this provides great flexibility, it also generates the problem of \emph{hidden state}, as the state of the notebook at a certain point may depend on more than the cells currently present in the notebook \cite{pimentelUnderstandingImprovingQuality2021,mackeFineGrainedLineageSafer2021,huangHowScientistsUse2025a}.
For instance, one can execute a cell that creates some variables, then delete the cell, and the variables are still defined (part of the \emph{hidden state}).
The flexibility provided by this per-cell execution model is often welcome, but it can also make notebooks harder to create and understand, and it is a leading cause of re-execution and reproducibility failure \cite{pimentelLargescaleStudyQuality2019}.
Marimo (for Python) \cite{Marimo2023,Gyarmatietal2025} and Pluto.jl (for Julia) \cite{Ritchie2022,Gevorkyanetal2023} implement, as \sys, a top-to-bottom execution semantics that eliminates the hidden state.
The methods used are different.
In Marimo, the absence of hidden state is achieved via a sophisticated analysis of the dependency between variables \cite{Gyarmatietal2025}.
Thus, Marimo can infer which cells need to be re-run when changes are made to ensure compliance with the semantics; when cells are deleted, the variables they introduced are also removed.
In \sys, we enforce adherence to the top-to-bottom semantics via our checkpointing kernel.
Our approach has some drawbacks, as we will discuss, such as an inability to cope with open files across cells.
Nevertheless, our approach is well-suited to support our verification tools.
It may be interesting, in future work, to explore the development of a variant of \sys that relies on a Marimo-based infrastructure for code execution.

\section{\sys: Syntax and Semantics} \label{sec-structure}

We begin by describing the syntactic \emph{structure} of a \sys instance (or notebook, in short) that highlights how it promotes the natural language cell descriptions while ensuring implementation stability, and its semantics or \emph{execution model} that enables validation via value inspections.

\subsection{Syntax}

Figure~\ref{fig-teaser} shows a \sys notebook in action.
A notebook consists of a \emph{top bar}, followed by
a sequence of \emph{comment}, \emph{action} and \emph{test}
cells.
The top bar, as shown in Figure~\ref{fig-teaser}, enables global
operations on the notebook, including running all its cells,
restarting from the beginning, and selecting the AI backend
used for code generation and verification.
Comment cells are identical to Jupyter markdown cells: they
are used to provide documentation or explanation within the
notebook.
Test cells are used to validate the values, and we will describe
them at length in Section~\ref{sec-testing}.

The bulk of the computational work of a notebook is carried out in action cells, much like Jupyter code cells.
However, the key difference is that the action cells of \sys promote the natural-language description of what the cell should do.
The code is generated automatically via AI, and it is stored to achieve implementation stability, but it is displayed only upon request.
Figure~\ref{fig-teaser} displays the first two actions cells of the \sys.
In the first, the  description instructs the AI to read a football dataset and display its top rows. 
In the second cell, the description asks to compute the number of matches each country played at home and away, and again display the first rows of the result. 

Users can \emph{focus} on a cell at a time by selecting it. 
Once a cell is focused, a series of buttons and options appear. 
Some buttons allow the user to \emph{edit} the
description and \emph{generate} the code from the description.
Other buttons allow \emph{clearing} the current code, so it can
be regenerated from a clean slate, and \emph{verifying} that the
code faithfully implements its description.
The buttons and their semantics will be described in more detail below
.

\sys notebooks are stored in Json, in the format used for Jupyter notebooks extended to accommodate descriptions, as well as testing and validation mechanisms.
For action cells and for the test cells discussed in the following, we store both the descriptions and the code in order to achieve implementation stability; obviously, we also store all outputs so that notebooks can convey results and information without the need for re-running them. 
We denote with $(d_1, c_1), \ldots, (d_n, c_n)$ the $n$ action cells of a notebook, where $d_i$ is the description of cell $i$, and $c_i$ is the code of cell $i$, for $1 \leq i \leq n$. 

\subsection{Semantics}\label{sec-semantics}

In a Jupyter notebook, cells can be executed in any order,
and the state of a notebook depends on the order of cell
execution, rather than on the position of cells in the
notebook \cite{pimentelLargescaleStudyQuality2019}.
For instance, if cells are deleted, their effects persist
in the notebook state, yielding what is known as the
\emph{hidden state} of Jupyter notebooks.
The flexibility afforded by arbitrary execution order can enable quick experimentation, and is useful when users can easily access the notebook variables, which carry the state. 
However, hidden state is a hindrance for natural language
users, who cannot easily keep track of the variables, and of how the code in the cells updates them. 

\sys is driven by language, and the cell descriptions constitute a narration of the computation.
In a narration, there is no notion that reading something twice changes its meaning.
That is, a user who cannot see the variable state more directly might be baffled by the fact that running the same description one more time changes the state of the notebook.

\mypara{Linear Execution: Cells and States}
Thus, we designed \sys to have a \emph{linear execution}
semantics, where the cells are executed in order, and each cell
is guaranteed to be executed in the state that results from the
\emph{previous} cell's execution.
Suppose that $c_1, \ldots, c_n$ is the code associated with the cells of a notebook.
Execution begins from the empty state $s_0$.
The code $c_i$ of each cell $i$ is executed in state $s_{i-1}$,
and produces state $s_i$.
This linear semantics ensures that each cell $i$
yields a \emph{unique} state $s_i$ that is precisely
the result of executing $c_1, \ldots, c_i$
starting from a blank slate $s_0$.
This linear semantics ensures that executing a cell
$i$ is idempotent: the resulting state $s_i$ is simply
the result of executing $c_1,\ldots,c_i$ in sequential
order.
In addition to eliminating hidden state, \sys's
linear semantics crucially allows the user to
associate a deterministic set of \emph{values}
$s_i$ to each cell $i$, which then lets them
validate those values, as described in Section~\ref{sec-testing}.

\mypara{The checkpointing kernel}
While a state $s_i$ \emph{can} be computed
simply by processing the notebook from
the beginning, in practice, that would
be terribly inefficient, as a user may
want to iteratively work on a single cell, and
it would be wasteful to rerun the entire
history on each iteration.
To support our linear semantics, we have implemented a \emph{checkpointing kernel} that enables state caching and makes state (re-)construction efficient (see Section~\ref{sec-implementation}).

\section{Validation and Testing}
\label{sec-testing}

\begin{figure*}[t]
  \centering
  \includegraphics[width=0.9\textwidth]{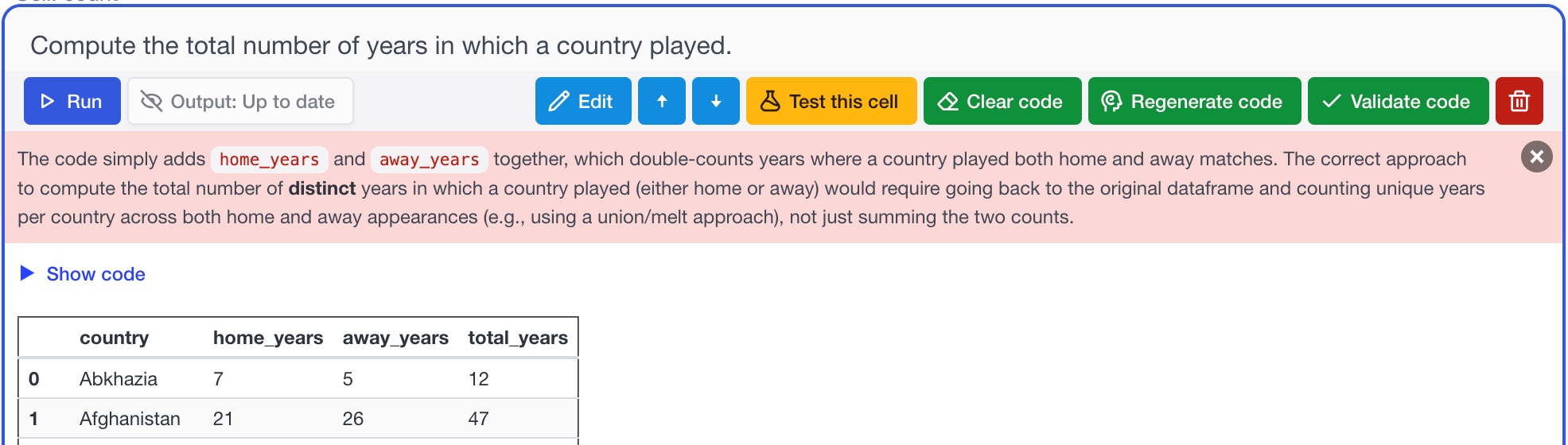}
  \caption{\sys can ask AI models to validate a cell's code against its description.  In this case, validation fails, and \sys displays diagnostic explanations returned by AI.}
  \Description{Screenshot of an action cell in \sys, where the cell validation failed.}
  \label{fig-failed-validation}
\end{figure*}

Jupyter notebooks are centered on code, with the assumption that users are able to understand the code they are writing, and can read the code present in notebooks that are shared with them.
However, when notebooks are created by, or shared with natural-language programmers, the code itself is inaccessible, and thus the key advantages of Jupyter notebooks --- verifiability, reproducibility, and extensibility --- are lost.
\sys preserves these benefits, by making the natural language descriptions, rather than the code, front and center.

Users describe computation using natural language, and rarely look at the code, evaluating it mostly from the results it produces.
This new usage mode raises two questions.
How can users be confident that the code implements their descriptions?
And, if the notebook has been authored by others and shared with them, how can they be sure that executing it is safe?

Our key observation is that while \sys users may not be able to evaluate code, they can understand data.
Thus, \sys provides methods for verifying implementation correctness based on AI, and on data inspection.
Next, we describe two mechanisms that \sys provides for \emph{validation} and \emph{testing}, which can be applied \emph{locally} to individual cells, or \emph{globally} across multiple cells.
%

\subsection{Validation}
\label{sec-cell-verification}

\mypara{Cell Validation}
By clicking on the \emph{Validate code} button of a particular cell, users can ask AI to verify that that cell's natural language specification is consistent with its code implementation.
To perform the validation, \sys sends all information used to generate the code to AI; the AI system prompt asks the AI model to reply \emph{Yes} if the code corresponds to the description, and to reply \emph{No} and provide diagnostic information otherwise.
If the validation result is negative, the user can click on \emph{Regenerate code}, and \sys will ask AI to regenerate the code, taking into account the problem diagnostics.
Figure~\ref{fig-failed-validation} illustrates an example in which validation fails to compute the total number of years in which a country has played. The diagnostic explains that the generated code is incorrect because of a ``double-counting'' error and what the correct approach should be.

\mypara{Safety via cross-validation}
Obviously, if one uses for verification the \emph{same} AI API used for code generation, the answer is very likely to be that the code is correct: the strength of the check lies in the ability to use a \emph{different} AI than the one that generated the code, avoiding dependence on a single AI source.
Indeed, it has been shown that AI models can contain hidden vulnerabilities, buried in the models in order to perform attacks \cite{hubingerSleeperAgentsTraining2024,Pearce2025,Siddiq2024}.
The attacks can be quite specific, triggered only when the code is generated in particular domains, or for specific applications, and within a range of target dates; the attacks can be quite hard to detect, as the AI API would perform normally for most users under most circumstances.
By allowing the cross-validation of code with respect to multiple independent AI APIs, \sys side-steps these vulnerabilities.

\mypara{Global Validation}
When a user opens a notebook shared by someone else, they can use the top-bar \emph{Verify} button (see Figure~\ref{fig-teaser}) and ask \sys to check that the entire notebook's cells correctly implement their descriptions.
Additionally, this process checks that the notebook as a whole does not contain any dangerous operations, defined as operations that can delete files, leak information, or alter the setup of the local host.
These validations are not perfect, of course, as code can be obfuscated.
Nevertheless, the ability to use an AI API of the recipient's choice in verification adds a layer of protection.

\subsection{Cell Testing}
\label{sec-unit-tests}

\begin{figure*}
  \centering
  \includegraphics[width=0.9\textwidth]{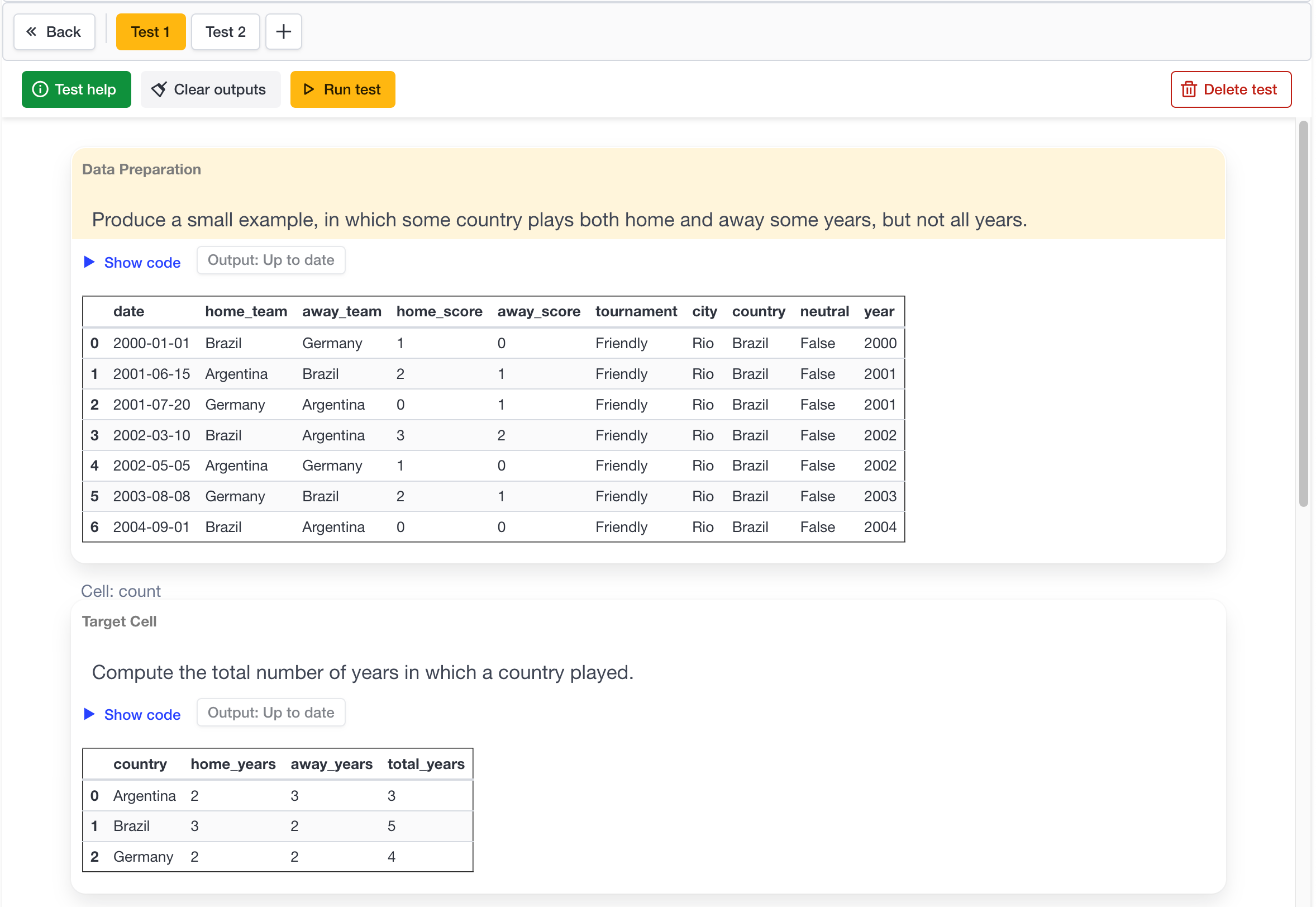}
  \caption{A cell test for verifying that the target cell of Figure~\ref{fig-failed-validation}, once fixed, correctly computes the number of unique years in which a country has played.  The Data Preparation cell sets up a simple example, enabling the user to assess the correctness of the target cell by inspecting its output.}
  \Description{Screenshot of an action cell in \sys.}
  \label{fig-years-unit-test}
\end{figure*}

While AI is a powerful tool for validating notebook correctness, it does not eliminate the need for a more direct verification method, in which the user directly assesses the correctness of the results.
Since the user may not understand code, we have built verification tools that are based on data, namely, \emph{cell tests}, which exploit \sys's linear execution semantics --- and lack of hidden state --- to enable users to systematically verify the values produced by the cells.

\begin{figure}
  \centering
  \includegraphics[width=0.9\columnwidth]{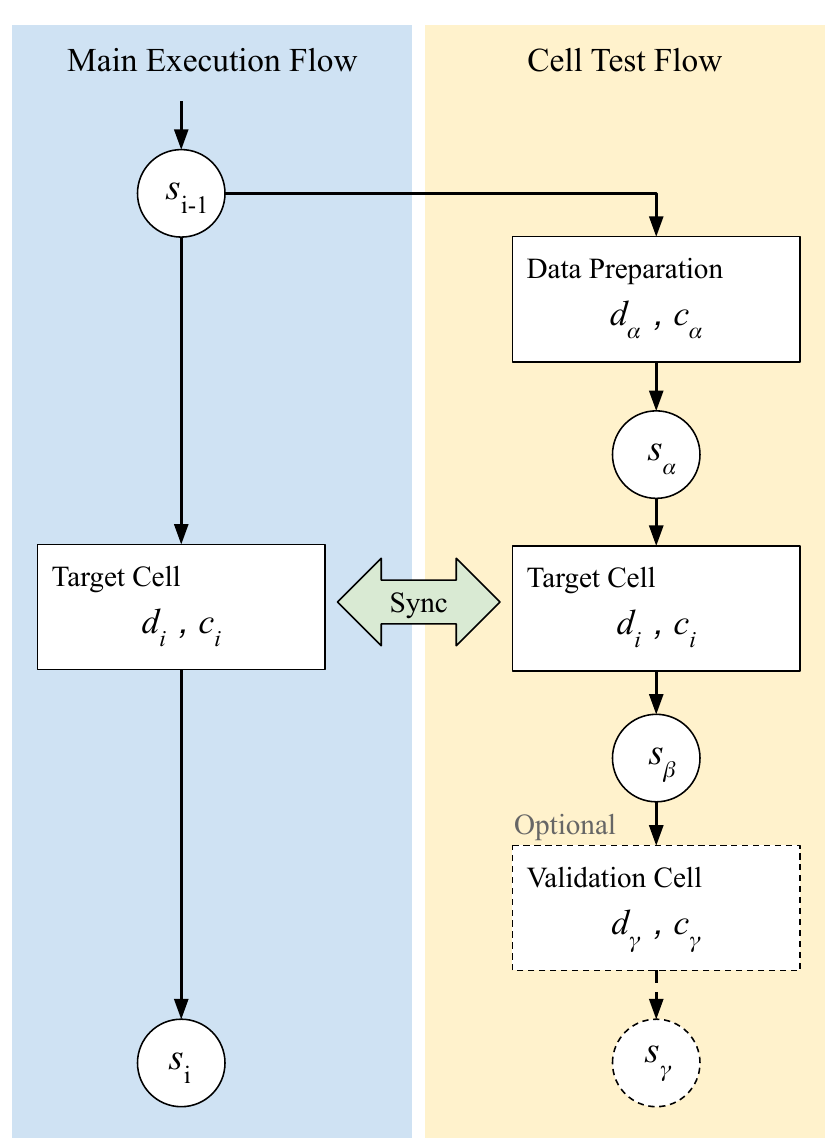}
  \caption{Execution flow of a cell test. To test a target cell $c_i$, \sys executes it in a parallel flow, where the data $s_{i-1}$ has been simplified into $s_\alpha$ by a data preparation cell.
  Users can check $s_\beta$ by inspection, or via a validation cell.}
  \Description{The figure depicts the execution flow for cell tests, as described in the article text.}
  \label{fig-test-flow}
\end{figure}

\mypara{Cell Test Flow}
The execution flow used to perform a cell test on a target cell $i$ is depicted in Figure~\ref{fig-test-flow}.
Recall (Section~\ref{sec-semantics}) that in the normal execution flow of the notebook, cell $i$ is executed in state $s_{i-1}$, which is the result of executing all cells from the beginning of the notebook up to $i-1$.
In the cell test flow, a \emph{data preparation} cell $\alpha$ is used to simplify some components of the state $s_{i-1}$, producing a new state $s_\alpha$ in which the code $c_i$ of the target cell is executed.
The user can then either directly examine the output, or write a validation cell (not used in the above Figure~\ref{fig-years-unit-test}) to check the resulting state $s_\beta$.

For example, recall the problem of counting the unique years in which a country played, and the failure of the cell validation in Figure~\ref{fig-failed-validation}.
A user can verify by inspection that the cell in Figure~\ref{fig-failed-validation} has been fixed by creating a \emph{cell test}, shown in Figure~\ref{fig-years-unit-test}
In the cell test, a \emph{Data Preparation} cell is used to produce simplified data, that sub-samples the full input data set according to some criteria specified in natural language.
The code of the \emph{target cell} of the test can now be run on this simplified data, and the user can verify by inspection that the output is correct, or specify a validation cell to check it.

\mypara{Testing without code modification}
\sys's linear semantics allow a user to test \emph{any} cell without any modification to the cell code.
As shown in the cell test flow in Figure~\ref{fig-test-flow},
the \emph{same} code of a cell (here, $c_i$) can be run in \emph{two} different environments: the regular environment (namely, $s_{i-1}$), and the test environment (namely, $s_\alpha$).
In contrast, with Jupyter notebooks, one can only test code that is \emph{wrapped} in functions or methods, so that it can be called under both regular and test settings, and consequently, much code ends up not being testable.

\mypara{Testing without data generation}
A second benefit of \sys's linear semantics is that users can leverage the execution model to create suitable test data for each cell.
In a typical unit test, a user is faced with the chore of cooking up appropriate test data from scratch.
Instead, in \sys, the user already has suitable test data, namely, the state $s_{i-1}$ in which the target cell normally runs (see Figure~\ref{fig-test-flow} again). All they need to do is simplify or modify some components in order to produce inpectable tests, which itself can be done via natural language in a \emph{data preparation} cell.

\mypara{Testing multiple functionalities}
Finally, a single target cell can have multiple tests associated with it. For instance, in Figure~\ref{fig-years-unit-test} there are two tests, as indicated in the tab bar; only the first one is displayed.
The tests allow users to test different functionalities of the target cell.
For instance, one or two tests can be used to test the cell under normal data, and further tests can indicate its behavior under pathological or missing data.

\subsection{Global Testing}
\label{sec-global-tests}

\begin{figure}
  \centering
  \includegraphics[width=\columnwidth]{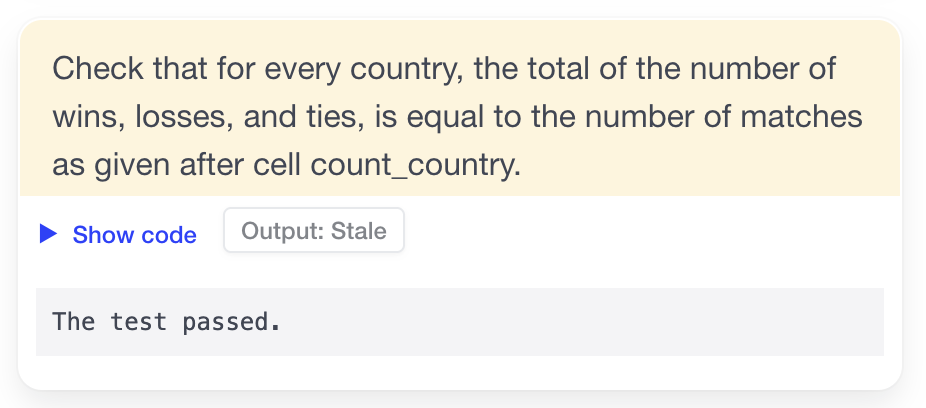}
  \caption{A global test. The test relates the state immediately preceding the test, with the notebook state after cell \texttt{count\_country}, which created a count of countries.
  The test checks that all the football matches that each country played are properly accounted.}
  \Description{A global test, checking that no countries are dropped from the computation from a reference cell, to the current cell.}
  \label{fig-global-test}
\end{figure}

\sys' linear execution model allow us to implement \emph{global tests}, which let users to specify assertions that relate the state of the notebook across \emph{different points} in the execution, to check that the overall computation, over the standard data, satisfies properties of interest.

\mypara{Specifying Global Tests}
Global tests crucially depend on \sys's linear semantics: specifically, that the result of executing the first $i$ cells always produces the state $s_i$.
To enable global tests, \sys automatically generates short nicknames for each such state $s_i$, which are then used to create \emph{namespaces} that hold the (checkpointed) state $s_i$.
Global tests can then use the cell nicknames to specify the states from which the variable values should be taken, thereby writing properties or assertions that relate the values of different states.
Figure~\ref{fig-global-test}, shows a global test that checks that all the countries defined after another reference cell nicknamed \texttt{cell\_country} are preserved at the current cell.

\mypara{Implementing Global Tests}
\sys executes global tests, by constructing a composite namespace in which all namespaces (states) of previous cells are included as sub-name-spaces.
In this composite namespace, the variable \texttt{x} of a cell with  \texttt{nickname} can be accessed as \texttt{\_\_state\_\_nickname.x}.
Thus, the composite namespace allows an assertion to \emph{simultaneously} access the states of \emph{all} previous cells.
The AI system instructions for global tests detail this access scheme, enabling AI to generate test code that contains the appropriate variable accesses.

In our experience, users mainly use global tests to check overall properties, and in particular, that data is appropriately preserved or discarded during filtering.

\section{Implementation} \label{sec-implementation}

\sys's key design goals: promoting the natural language computation descriptions, and enabling value validation, rely crucially upon two abilities.
First, the AI needs to be able to use a cell's description, and other context information available from previous computation, to \emph{generate the code} that faithfully implements the user's natural language description for that cell.
Second, we need to be able to execute the generated code and 
faithfully implement \sys's linear execution semantics.
In Sections~\ref{sec-generation} and~\ref{sec-execution} respectively, we describe how \sys implements these two key capabilities.

\mypara{Python as the Foundation}
Python serves as the foundational language
for \sys, powering both its code generation
and execution engines.
This choice has three motivations.
First, AI is quite proficient at generating Python,
as training data is very abundant, including in the
data science realm.
Second, Python is already in use in (Jupyter) notebooks,
and has widely-used libraries for data science and
visualization.
Third, the meta-programming and dynamic execution
capabilities of Python make it easy to implement
the kernel and testing harnesses we built into \sys,
as we shall describe below. 

\subsection{Generation} \label{sec-generation}

\sys generates code for each cell
by querying an AI API with a combination of
\emph{global} system instructions, and a
\emph{local} (cell-specific) prompt.
(Currently, \sys supports Claude Code or Gemini,
but other AI APIs can be added easily).
%

\mypara{Local Prompt}
The prompt for generating the code $c_i$ for a cell $i$
was determined via experimentation, and includes:
\begin{itemize}
    \item The \textbf{description} $d_i$ of the cell to be generated.  The AI is instructed to return code that implements this description.
    \item If a previous code implementation \textbf{code} $c_i$ for cell $i$ already exists, we include it in the request, asking the AI to revise the code, if necessary, to implement $d_i$. The AI request is phrased so that the code is revised only if the current implementation does not implement the description. 
    \item The \textbf{descriptions} $d_j$ and \textbf{code} $c_j$ of all previous cells $1 \leq j < i$.  This creates context for the AI, so that it can understand the previous work done in the notebook, and the variables and modules that are in the scope.
    \item The \textbf{variables} defined in the state $s_{i-1}$ in which the cell will be executed.  We provide to the API a list of variables, each with type information (but not value information). For Pandas dataframes, we include information on the names and types the columns, so that the AI can understand how to operate on the dataframe.
    \item Optionally, the \textbf{output} of the previous cell $c_{i-1}$.  This is useful to allow descriptions to refer to that output as well; for instance, one can write ``increase that value by 10\%, and select all data where the price is above that threshold''.
\end{itemize}

\mypara{Global System Instructions}
In addition to the above, some global
contextual information regarding files
and domain-specific instructions are
added to each AI request:
\begin{itemize}
    \item \textbf{Files.} An AI API does not
    have access to the user's filesystem, and
    cannot know where files of interest are.
    To allow for file access, \sys includes
    a \emph{files tab} where users can select the
    locations of files used in the notebook,
    such as datasets.
    In this way, the AI knows which paths to
    include in the code.

    \item \textbf{Instructions.}  Some projects have
    domain-specific background knowledge that is useful
    for AI generation: for example, methods for accessing data,
    other APIs available, and so forth.
    \sys includes a \emph{instructions tab} where users
    can specify such background information.
\end{itemize}
For example, Figure~\ref{fig-teaser} shows the beginning a notebook to analyze a football (soccer) dataset.
The \emph{Files} tab specifies the location of the \texttt{football.csv} file in the user machine's filesystem.
The AI API (here, Claude Haiku) is thus made aware
of the location of files that the user might mention,
and is able to translate the cell's description into
code that loads the dataset.

\mypara{Privacy and Cost Considerations}
Including cell output in the AI input can pose a privacy risk, as cell output can include sensitive information such as customer identities or data.
For this reason, \sys enables to toggle on and off the sharing of cell outputs with AI on a per-notebook basis.
In any case, to reduce AI cost, we do not include graphics such as plots and figures in the information relied to AI.
We do not include the value of the variables (beyond the names of dataframe columns) in the information passed to AI.
Similarly to outputs, including these values incurs the risk of leaking confidential information.
Further, having as context the code $c_0, \ldots, c_{i-1}$ that generated the variables seems sufficient in practice.

\subsection{Execution}  \label{sec-checkpointing-kernel} \label{sec-execution}

\sys's linear execution semantics are the key to enabling
hidden state and associating a deterministic values that
can be validated of each cell and across multiple cells
(Section~\ref{sec-testing})
However, a naive implementation of the linear execution
semantics, which, simply re-runs the notebook from the
beginning each time a cell is executed, would be terribly
inefficient.

A better approach could be to repurpose the Jupyter kernel,
where cells are executed in the current kernel state: the
resulting state becoming the new current state.
To implement \sys's linear semantics, specifically,
to execute cell $i$, we could check the index of the
last executed cell: if it was $i-1$, we could simply
execute cell $i$; otherwise, we restart the kernel and
execute cells $1, 2, \ldots, i$.
Sadly, while better than the naive implementation,
this repurposed kernel would still be rather inefficient.
In particular, users often refine a particular
cell, until they are satisfied with its behavior.
Re-running the notebook each time from the beginning
would make this common use case most inefficient.

\mypara{A Checkpointing Kernel}
To implement our linear semantics efficiently,
we have developed a \emph{checkpointing kernel},
which stores the states obtained after executing
each cell in the proper order.
Recall that in our linear model, execution starts
the empty state $s_0$, and each cell $c_i$ is executed
in state $s_{i-1}$ to produce state $s_i$.
With checkpointing, if cell $c_i$ is executed
multiple times, for instance, as the user is
iteratively refining their prompt, only the
code $c_i$ needs to be run, producing each
time new states $s_i$.

\mypara{Checkpoints as Python Namespaces}
The checkpointing kernel stores states
as Python namespaces.
This induces one limitation: the states cannot
track \emph{external} state, such as the state
of open files.
Fortunately, our typical users have the same
limitation.
Users who perform data science in natural
language typically read and write data within
the scope of individual cells, rather than
carrying open file descriptors across multiple cells.
To support the need to restart execution
(for example, to re-read external files that changed),
\sys includes top-bar button to \emph{clear}
all kernel states, ensuring a full restart.

\mypara{Implementing the \sys Kernel}
We have implemented \sys as a web application
that runs from localhost, similarly to the
classic \texttt{jupyter notebook} module.
The checkpointing kernel is implemented
as a stand-alone Python process, communicating
with \sys via HTTP; a kernel is created
automatically for each notebook.
Relying on HTTP(s) makes it possible to run the
kernel on a different machine from where
the UI is located, even though we are not
using this capability in the current
version of \sys.

\subsection{Orchestration} \label{sec-orchestration}

To efficiently and interactively implement
\sys's linear execution semantics, the \sys
kernel needs to orchestrate the re-generation
and re-execution of cells, depending on the
specific operations performed on the notebook.
Orchestration is tricky because (recall, from Section~\ref{sec-generation})
the prompt for generating the code for cell
$c_i$ includes descriptions of the \emph{variables}
defined in the state $s_{i-1}$ which provide
important context for the code generation task.

Suppose, for example, that a user generates
code for cells $1$ through $10$, and then goes
back and \emph{changes} the description of cell
$4$, so that now perhaps also the content of a
variable has changed meaning.
At this point, the \sys kernel must also mark
the code for cells $4$ through $10$ as invalid.
At the same time, the kernel need not
\emph{regenerate} those cells eagerly.
Perhaps the user wants to \emph{iterate}
applying other changes to cell $4$.
Nevertheless, the kernel must remember
that \emph{eventually}, if the user wants
the values for the cells $4-10$, then the
code and values for those cells must
be regenerated.

\sys implements orchestration by using \emph{indices} to
track cell validity, and then using the indices
to compute the minimal set of cells that need
to be generated or executed to efficiently
implement \sys's linear semantics for any
operation performed on the notebook.

\mypara{Kernel State: Indices}
To determine the cells that need code regeneration or execution, \sys tracks the indices of the last action cell:= 
\begin{itemize}
\item whose \emph{code} is valid: $\codeidx$,
\item whose \emph{output} is valid: $\outidx$, and
\item whose code was \emph{executed}: $\execidx$.
\end{itemize}
Code and output are preserved in the stored notebook, making it possible to load a notebook and display it complete with its results, as usual.
Correspondingly, the indices $\codeidx$ and $\outidx$ are stored with the notebook.
The \sys kernel maintains the invariant that
\begin{equation} \label{eq-code-geq-out}
  \codeidx \geq \outidx
\end{equation}
which captures the intuition that valid output
can only be generated by valid code.

Conversely, the states of the execution kernel are not stored; rather, when a notebook is read, the states must be re-created by rerunning the notebook from the beginning. 
Consequently, the execution index $\execidx$ is not stored with the notebook, and is set to~0 when a notebook is read or created.

\mypara{Kernel State: Effects of Operations}
Next, we describe how the kernel \emph{effects}
the indices according to the operations performed
on the notebook.
Certain operations like code generation and
execution each have a \emph{precondition} on
the indices that must be true before the
operation can be performed.
As we explain shortly, the kernel recursively
re-generates or re-executes cells as needed,
to ensure that the preconditions for the
requested operation are satisfied.

\paragraph{Reading the notebook.}
When the notebook is read, \sys sets $\execidx := 0$,
and we leave $\codeidx$ and $\outidx$ unchanged from
the values they have in the stored notebook.
This allows users to read a notebook, whether produced
by themselves or shared with them, and still consider
the code and output valid, unless the notebook is
further modified.
Note that a notebook is fully verifiable, as both
the associations of descriptions and code, and the
associations of code and output can be checked at
any time, as discussed in Section~\ref{sec-testing}.

\paragraph{Editing the notebook.}
When the description of cell $i$ is edited, or when
cell $i$ is deleted or inserted, or when cells $i$
and $i+1$ are swapped in the notebook,
\sys invalidates the code and output
beyond position $i$:
\begin{align*}
    & \mathrm{Effect:} & \codeidx & := \min\set{\codeidx, i-1} \\
    &                  & \outidx  & := \min\set{\codeidx, i-1}
\end{align*}

\paragraph{Code generation.}
Code can be generated for a cell $i$ when $\outidx \geq i - 1$,
so that the information on the variable context for $c_i$ is available.
If code can be generated, then the code $c_i$ becomes valid,
and the outputs for any cell beyond $i-1$ become stale,
as $c_i$ still needs to be run.
\begin{align*}
    & \mathrm{Precondition:} & \outidx & \geq i - 1 \\[2ex]
    & \mathrm{Effect:}       & \codeidx & := i \\
    &                        & \outidx & := \min\set{\outidx, i-1}
\end{align*}

\paragraph{Code execution.}
Code can be executed if it is valid,
and if the output from the previous
cell is valid, indicating that the
state $s_{i-1}$ in which to execute
$c_i$ is present.
Note that executing a cell does not
cause the output of subsequent cells
to become stale.
The conditions for executing cell $i$ are:
\begin{align*}
    & \mathrm{Precondition:} & \codeidx & \geq i \\
    &                        & \execidx & \geq i - 1 \\
    &                        & \outidx & \geq i - 1 \\[2ex]
    & \mathrm{Effect:} & \outidx & := \max\set{i, \outidx} \\
    &                  & \execidx & := \max\set{i, \execidx}
\end{align*}
Execution in \sys is meant to be idempotent,
and for this reason, if $\outidx \geq i$,
when the user calls for the execution of
cell $i$, the kernel is not used; rather,
we simply use the cached state $s_i$ and
the cached output of the previous execution.

\mypara{Orchestration using Kernel Indices}
Recall, from Section~\ref{sec-structure} and Figure~\ref{fig-teaser}, that \sys
has a variety of buttons that enable the user to generate (or regenerate)
the code for a single cell and run either a single cell or the entire notebook.

When one of the buttons is pressed, calling for execution or code generation, \sys invokes an orchestration engine that computes the minimum set of code generation and execution that is needed to comply with the request, and performs the sequence of generations and executions.
For instance, assume that:
\begin{itemize}
    \item A notebook with $n$ cells is read, so that $\codeidx = \outidx = n$,and $\execidx = 0$.
    \item The description of cell $i$ is edited.
    \item The execution of cell $k$, for $i < k \leq n$, is requested by clicking on the Run button of cell $k$.
\end{itemize}
Then, the orchestration engine would, in order:
\begin{itemize}
    \item Run cells $1, \ldots, i-1$ to generate the context $s_{i-1}$ and the output and variables needed for generating the code $c_i$ of $i$;
    \item Generate the code $c_i$ for cell $i$, and run it;
    \item For each cell $j \in \set{i+1, \ldots, k}$, in order, first generate the code $c_j$ using the context of $j-1$, and then run it.
\end{itemize}
The orchestration engine is implemented
in the Javascript front-end of \sys, so
that the sequences of code generation and
execution can be easily interrupted if desired.
The back-end, driving the kernel execution,
checks the preconditions and implements the
effects of all notebook operations, thus
ensuring the consistency of the notebook state.

\section{Discussion}

We put \sys to the test by implementing the main data analysis tasks in the university committee.

\subsection{Strengths}

This exercise was successful.
Most often, individual analysis steps were simple, and they could be described via short prompts that generated obviously correct results.

\mypara{Iterative}
For complex tasks, the ability to perform \emph{cell tests} to verify the
computation performed by cells was essential.
In these more complex cases, we proceeded in an iterative fashion, refining a cell's prompt until the results seemed correct.
The linear execution semantics and checkpointing kernel were helpful in such an iterative approach, as we could focus on perfecting the current cell, without having to worry about restarting the execution from scratch to avoid carrying over the bad state from the previous cell version.
Once the cell seemed to work, we could use \emph{cell tests} to try it on smaller data and verify by inspection that it behaved correctly.
Generating simplified test data via natural language was quick and effective.
The example reported in Figure~\ref{fig-years-unit-test} was typical: in cell tests, we sought simple data, where the results could be checked by inspection; we often specified some aspects of the simplified data, to test  corner cases.

\mypara{Accessible}
\sys notebooks were usable by those with little coding experience.
Even among those familiar with code, they became a favorite for an unexpected reason: they were often \emph{faster} to use than Jupyter notebooks augmented with AI.
In \sys, one needed only to type a prompt (e.g., ``group the students by major, and plot their graduation time'') and click \emph{Save and Run} (or press shift+Enter).
With Colab or VSCode, one needed more clicks to call up the AI helper, then generate the code, accept it, and run it.

\mypara{Collaborative}
When performing data analysis collaboratively in the haste of a meeting, this speed difference mattered.
It also mattered that any participant could suggest a prompt to be run (whereas only a few would have been able to suggest code).
The ability to iterate on a cell without having to worry about rerunning the notebook from the beginning (the need for which is quite unclear to the uninitiated) was also a positive.

\subsection{Limitations and Future Work}

\mypara{Configuring API Keys}
The primary barrier to adopting \sys was the requirement for AI API keys.
While the \sys settings provide direct links to the necessary provider pages, the process of securing keys and configuring billing accounts remained a significant hurdle for users.

\mypara{External State}
The chief limitation of the checkpointing kernel is its inability to track external (to Python) state.
In practice, users tend not to leave files open across notebook cells, as the concept of a ``file descriptor'' is not a natural one for non-programmers.
However, database access is more problematic.
Once connected to a database, if a cell with the prompt ``add to the database a student with a major in Biology'' is executed twice, the result is not idempotent: two students are added.
It remains future work to extend the linear semantics to databases via transactions and logs.
So far, \sys is amenable to analysis tasks where data can be read, and written, in one fell swoop that fits in a single cell.
Fortunately, many data science tasks fall in this category.

\mypara{Code Regeneration}
In settings where \sys can be used, there are still rough spots in usability, which we plan to address in future work.
Most significantly, when a user modifies the prompt of one of the early cells in the notebook, \sys regenerates the code and runs all subsequent cells (see Section~\ref{sec-orchestration}).
The code is regenerated because if we change the code $c_i$ for a cell $i$, the variables available for the execution of cell $i+1$ (and more subtly, the \emph{meaning} of the variables available) may change.
Therefore, to ensure correctness, we regenerate the code of  cell $i+1$ before executing it, so that the changes in variable availability and meaning can be accounted for.
Regenerating the code of all cells that follow a modified cell is both slow and overly conservative: typically, the code of most, if not all, such cells is still valid.
In future work, we will seek to limit code regeneration by asking AI to first identify which cells need their code regenerated.

\section*{Code Availability}
\sys is an open source project~\cite{dealfaroPlainbook}.
The code of both \sys and the checkpointing kernel that it uses is released under the BSD 3-clause license. 
\sys can be installed using the pypi \texttt{pip} package manager. 

\bibliographystyle{ACM-Reference-Format}
\bibliography{extra}

\end{document}